\documentclass[aps]{revtex4}
\usepackage{graphicx}
\usepackage[all]{xy}
\usepackage{amsmath}
\usepackage{amssymb}
\usepackage{color}

\newcommand{\bb}{\bibitem}
\newcommand{\bes}{\begin{subequations}}
\newcommand{\ees}{\end{subequations}}
\newcommand{\benn}{\begin{eqnarray*}}
\newcommand{\eenn}{\end{eqnarray*}}

\newcommand{\be}{\begin{equation}}
\newcommand{\ee}{\end{equation}}
\newcommand{\ben}{\begin{eqnarray}}
\newcommand{\een}{\end{eqnarray}}

\def\ben{\begin{eqnarray}}
\def\een{\end{eqnarray}}
\def\be{\begin{equation}}
\def\ee{\end{equation}}

\begin{document}

\title{Localized four-dimensional gravity in the D-brane background with NS $B$ field} 

\author{ R.C. Fonseca}
\affiliation{Departamento de F\'\i sica, Universidade Estadual da Para\'\i ba,\\ Centro de Ci\^encias Exatas e Sociais Aplicadas - CCEA, Rua Bara\'unas, 351 - Bairro Universit\'ario,\\ 58429-500, Campina Grande, Para\'\i ba,Brazil}

\author{F.A. Brito}
\affiliation{Departamento de F\'\i sica, Universidade Federal de Campina
Grande, \\Caixa Postal 10071, 58109-970 Campina Grande, Para\'\i ba, Brazil}

\author{L. Losano}
\affiliation{Departamento de F\'\i sica, Universidade Federal da
Para\'\i ba, \\Caixa Postal 5008, 58051-970 Jo\~ao Pessoa, Para\'\i ba, Brazil}

\date{\today}

\begin{abstract}
We calculate small correction terms to gravitational potential near $p$-branes embedded in a constant NS $B$ field background in the context of M-theory or string theory. The normalizable wave functions of gravity fluctuations around the brane describe only massive modes. We compute such wave functions analytically. We estimate the correction to gravitational potential for small and long distances, and show that there is an intermediate range of distances in which we can identify $4D$ gravity on the brane below a crossover scale given in terms of components of the $B$ field. The $4D$ gravity is metastable and for distances much larger than the crossover scale the $5D$ gravity is recovered.
\end{abstract}

\maketitle

%%%%%%%%%%%%%%%%%%%%%%%%%%%%%%%%%%%%%%%%%%%%%%%%%%%%%%%%%%%%%%%%%%%%%%%%%%%%%%%%%%%%%%%%%

\section{Introduction}

In the original idea of Randall and Sundrum (RS) \cite{RS1} scenario, the five-dimensional gravity is coupled to a negative cosmological constant and a $3-$brane sourced by a delta function. The solution in such setup is a symmetric solution given in terms of two copies of AdS$_5$ spaces patched together along the $3-$brane. Although in this setup the fifth dimension is infinite the volume of the $5D$ bulk space is finite because the geometry is warped. As a consequence this allows the emergence of graviton zero mode responsible for $4D$ gravity on the brane. This is not necessary true for spaces whose volume diverges, because no such zero mode emerges anymore. This was first shown by Gregory-Rubakov-Sibiryakov (GRS) \cite{RGS} and Dvali-Gabadadze-Porrati (DGP) \cite{DGP}. The interesting consequence of such an alternative setup is that $4D$ gravity on the $3-$brane now emerges due to massive modes and then gravity is metastable. However, gravity massive modes can live long enough before escaping from the $3-$brane to produce $4D$ gravity within a sufficiently large scale $r_c$, the crossover scale. Two recent discussions on this matter involving thick branes, scalar fields and supergravity can be found, {\it e.g.}, in \cite{FBL1, FBL2}. 

In the present study we investigate such a scenario in braneworld models which arise from a sphere reduction in $M-$theory or string theory, as the near horizon of $p-$branes with a constant background $B$ field on the worldvolume, whose dual is a non-commutative Yang-Mills field theory \cite{PO1}. Thus, based on the correspondence between gauge theories and string theory in curved backgrounds, it is possible to investigate some important aspects of noncommutative gauge theories considering gravity solutions with $B$ fields. Such solutions provide dual descriptions of non-commutative field theory which allow to analyze the phase structure and the corresponding validity of the different descriptions. One can calculate the two-points correlation function involving components of momentum in the direction of the $B$ field \cite{PO1,PO2}. In \cite{PO3}, were found that the examples where gravity-trapping emerges can all be lifted back to become
the near-horizon regions of M-branes or D$p-$branes with $p \leq 5$. These are precisely the branes for which a natural gravity-decoupling limit exists, which is an indispensable condition for the possibility of establishing a Domain-wall/QFT correspondence. 

However, we shall consider string theoretic $p-$brane solutions, for the $p > 5$ case, to which the space of the extra large dimension has infinite volume, where the DGP and GRS scenarios takes place, since as we shall show, this produces massive graviton modes.
We found a number of interesting results in this context. From the graviton wave equation, in the background of $5D$ domain-walls that originate in string theory as $p-$branes in the constant background $B$ fields, we find that the massless graviton wave function is not normalizable for nonzero background $B$ fields, indicating that the massless graviton is not localized on the brane, however, there are massive gravitons localized on the brane. In $p=6$ case, as previously emphasized, we shall focus on $4D$ metastable gravity, that is, the fact that  gravity becomes four-dimensional for distances very much smaller than the crossover scale and five-dimensional gravity for distances very much larger than such scale. In doing so, we shall find the Newtonian potential induced by the gravity massive modes of a Schroedinger-like equation for the gravity fluctuations around the brane solution. 

The paper is organized as follows. In Sec.~\ref{gw} we introduce the graviton wave equation and apply to a general $p-$brane solution. We explore three examples. In Sec.~\ref{conclu} we make our final discussions.

\section{The Graviton wave equation}\label{gw}
%\subsection{The background $B-$field}
In this section, will develop a general graviton wave equation in the background of five-dimensional domain-walls that originate in string theory as $p-$branes in constant background $B$ (we do not address explicitly the cases for $E$ fields here). Consider the metric of a $p-$brane expressed in the string frame in a constant NS $B$ field in the 2,3 directions \cite{PO1,PO2}
\ben\label{P.1}
ds^2_{10}&=&H^{-1/2}\Big(-dx_0^2+dx_1^2+h\,(dx^2_2+dx^2_3)+dy^2_i\Big)+l^4_s\,H^{1/2}\Big(du^2+u^2d\Omega^2_{8-p}\Big),
\een where
\be\label{P.2}
H=1+\frac{R^{7-p}}{l^4_s\,u^{7-p}},\quad\quad\quad \frac{1}{h}=\frac1H{\rm sin}^2\theta+{\rm cos}^2\theta,\nonumber
\ee $u=r/l^2_s$ is a dimensionless radial parameter and $i=1,..,p-3$. The $B$ field and dilaton are given by
\ben\label{P.3}
B_{23}=\frac{h}{H} \,{\rm tan}{\,\theta},\quad\quad {e}^{2\,\varphi}=h\,g_s^2\,H^{\frac{3-p}{2}}\\
{\rm cos}\theta\,R^{7-p}=\Gamma\left(\frac{7-p}{2}\right)g_sl_s^{p-3}\,N(4\pi)^{\frac{5-p}{2}}
\een where $N$ is the number of $p$-branes and $g_s$ is the asymptotic value of the coupling constant. The non-commutativity parameter $\theta$ is related to the asymptotic value of the $B$ field with $B_{23}^{\infty}={\rm tan}\theta$  \cite{PO1}.%and $g\equiv g_\infty$ is the asymptotic value of the coupling constant \cite{PO1}.

Let us now, consider the following decoupling limit, in which the $B$ field goes to infinity. The rescaling of the parameters should be taken as in the following
\ben\label{P.3.1}
l_s\rightarrow0,\quad l_s^2{\rm tan}\theta=b,\quad \overline{x}_{2,3}=\frac{b}{l_s^2}\,x_{2,3},
\een where $b,u,\overline{x}_{_{\lambda}}$ and $g_sl_s^{p-5}$ are maintained fixed. Thus,
\ben\label{P.3.2}
\frac{ds^2}{l_s^2}=\left(\frac{u}{R}\right)^{\frac{7-p}{2}}\Big(-dx_0^2+dx_1^2+\overline{h}\,(d\overline{x}^2_2+d\overline{x}^2_3)+dy^2_i\Big)
+\left(\frac{R}{u}\right)^{\frac{7-p}{2}}
\Big(du^2+u^2\,d\Omega^2_{8-p}\Big),
\een with  $\overline{h}\,\,^{-1}=1+b^2\,(u/R)^{7-p}$.

One can see that for a D3-brane solution, i.e., $p=3$, the metric (\ref{P.3.2}) as  $u\to0$ describes the geometry of the $AdS_5\times S^5$ spacetime. It was conjectured in \cite{PO1} that this is the gravity dual of the Yang-Mills theory with noncommuting 2,3 coordinates.

For $p\neq5$, we consider the change of coordinates to $u/R\equiv(1+k|z|)^{2/(p-5)}$ (suggested in Ref. \cite{PO3}, which opens the possibility of evaluating cases related to the sign of $k$), which leads us to
\ben\label{p.5}
\frac{ds^2}{l_s^2}=(1+k|z|)^{\frac{7-p}{p-5}}\left(-dx_0^2+dx_1^2+\overline{h}\,(d{x}^2_2+d{x}^2_3)+dy^2_i+dz^2\right)
+(1+k|z|)^{\frac{p-3}{p-5}}d\Omega^2_{8-p},
\een
where we have dropped the ($-$) on the coordinates.

We shall now address the issue of four-dimensional gravity localization on the remaining {\it four} worldvolume coordinates after considering that $y_i$ are wrapped around a compact $(p-3)$-dimensional manifold.

Recalling that the metric is expressed in string frame, the equation of motion for the graviton fluctuation $\Phi=g^{00}\,h_{01}$ is given by
\be\label{flut}
\partial_M\sqrt{-g}\,{e}^{-2\,\varphi}\,g^{MN}\,\partial_N\,\Phi=0,
\ee where $h_{01}$ is associated with the energy-momentum tensor component $T_{01}$ of the Yang-Mills theory. For $p\neq5$, consider the ansatz \cite{PO0}
\be\label{an1}
\phi(x,z)=\,\phi(z)\,{\rm e}^{i\,p.x}=(1+k|z|)^{\frac{9-p}{2\,(5-p)}}\,\psi(z)\,{\rm e}^{i\,p.x}.
\ee
Thus, we obtain a Schroedinger-like equation
\be\label{sch.1}
-\frac{d^2}{d\,z^2}\,\psi(z)+U(z)\,\psi(z)=m^2\,\psi(z),
\ee where we have the potential
\be\label{pot.1}
U(z)=\frac{(p-9)(3\,p-19)\,k^2}{4\,(5-p)^2(1+k|z|)^2}+\frac{(p-9)}{(5-p)}\,k\,\delta(z)+\frac{\alpha^2}{(1+k)^{2(7-p)}(5-p)},
\ee with $\alpha\equiv b\sqrt{p_2^2+p_3^2}$ and $m^2=- p_\mu p^\mu$, where $\mu=0,1,...,p$ is a worldvolume index. The cases studied in this paper are focused on D6-branes, i.e., $p=6$ and the general form for the potential \eqref{pot.1} now reduces to the following simplified form 
\be\label{GPot}
U(z)=\frac{\nu}{\left(1/k+|z|\right)^2}+3\,k\,\delta(z),\quad\quad \nu=\frac34-\frac{\alpha^2}{k^2}.
\ee  The wave function solution can be written in general form
\be\label{Be1}
\psi_m(z)=C_{m}\,\sqrt{\frac{1}{k}+|z|}\,
\left[ J_{\nu}\left( m\left(\frac{1}{k}+|z|\right)\right)-F(m)\,N_{\nu}\left( m\left(\frac{1}{k}+|z|\right)\right)
\right]
\ee where $J_{\nu}$ and $N_{\nu}$ are Bessel functions of the first and second kind, respectively and $m$ are KK massive modes. Since we are interested in the correction terms to the four-dimensional Newton law
between two unit masses on the brane then we repeat the procedure in this situation. It is necessary to obtain the probability of
gravity with KK-modes on the brane. The asymptotic behavior of $|\psi\left(0\right)|^2$ depends on the magnitude of the argument in the Bessel functions and the
normalization factor $C_{m}$ and $F(m)$ certain conditions, i.e.
\be\label{Nor}
|C_{m}|^{2}=\frac{m}{2\,\left(F^{2}(m)+1\right)},
\ee
\be\label{F1}
F \left( m \right) = \frac{  \left( 4\,k+2\,\nu\,k\right) J_{{\nu}}
 \left( {\frac {m}{k}} \right) -{2\,m}\,J_{{\nu+1}} \left( {\frac {m}{
k}} \right)  }{    \left(4\,k+2\,\nu\,k\right) N_{{\nu}} \left( {\frac {m}{k}} \right) -{2\,m}\,N_{{\nu+1}}
 \left( {\frac {m}{k}} \right)  }.
 \ee

In such a regime, it is reasonable to
approximate the potential generated by discrete massive graviton states as a summation of Yukawa-like potentials, which makes the
total effective potential to have the form \cite{SDW,MM,SCH} 
\be\label{pot}
V (r) = \psi^2_0(0)\frac{{\rm{e}}^{-m_0\,r}}{r}+\sum_{n}{\psi^2_n(0)}\frac{{\rm{e}}^{-m\,r}}{r}=V_0+\delta\,V(r),
\ee 
where the first term is contribution of zero mode and the second term corresponds to
the correction term which is generated by the exchange of KK-modes. Now, using the Lommel's formula, $J_{\nu+1}(x)\,N_{\nu}(x)-N_{\nu+1}(x)\,J_{\nu}(x)=2/\pi\,x$, the asymptotic forms of Bessel function, $J_{{\nu}} \left( x \right) \sim\sqrt {{\frac {2}{x\pi }}}\cos
 \left( x-1/2\,\nu\,\pi +1/4\,\pi  \right)
$ and $J_{{\nu}} \left( x \right) \sim\sqrt {{\frac {2}{x\pi }}}\sin
 \left( x-1/2\,\nu\,\pi +1/4\,\pi  \right)$ for $x>>1$, $J_{{\nu}} \left( x \right) \sim \left( 1/2\,x \right) ^{\nu}
$, $ N_{{\nu}} \left( x \right) \sim-{ {1/\pi } \left( 1/2\,x \right) ^{-\nu}}
$ for $x<<1$, the probability density assumes the simplified forms (and independent of the sign of k):
\ben\label{Prob1}
\left| \psi_m \left( 0 \right)  \right|^2&\sim&\frac{4\,{m}^{2}}{{\pi }\, \left( \varepsilon^2+ {4\,{m}^{2}} \right)},\quad\quad \quad\quad \quad\quad \quad\quad 
 \quad\quad\quad\,\quad\,\,\,\,\quad\quad m>>|k|;\\
 \left| \psi_m \left( 0 \right)  \right| ^{2}&\sim&  \frac{2\,m }{|k|\, \left( \nu+1
 \right) ^{2}}\,\left( {\frac {m}{|k|}} \right) ^{2\,\nu}
  ,\quad\,\,\,\,\,\,\,\quad\quad\quad\quad\,\quad\,\,\,\,\,\quad \quad\quad\quad\,\, m<<|k|;
\een where $\varepsilon^2=k^2\,(4+2\,\nu)^2$ The choice of the sign of $k$ has to be correlated with the
matching condition across the singular domain wall source.

\subsection{General forms for the gravitational potentials in $p=6$ case}

\subsubsection{First case: ``Box''$-$like potential}
\begin{figure}[ht]
\includegraphics[{height=05cm,width=06.5cm,angle=00}]{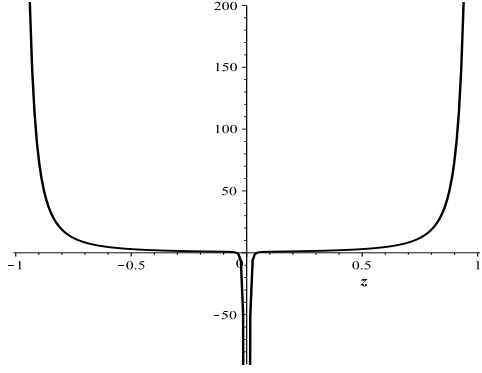}
\caption{Profile of the potential (\ref{GPot}) as a function of $z$ for $k<0$ with $\nu<<1$.}\label{fig.1}
\end{figure}

In the first case, which relates to $k <0$, the problem is similar to that of a particle trapped into an infinite box up to a delta function that we use to impose boundary conditions --- See Fig.~\ref{fig.1} --- which massive modes solution of the Schroedinger-like problem is given by (\ref{Be1}), taking into account the adjustment of parameters.
%%%%%%%%%%%%%%%%%%%%%%%%%%%%%%%%%%%%%%%%%%%%%%%%%%%%%%%%%%%%%%%%%%
Our analysis takes place in two distinct regions, where we obtain the probabilities for existence of gravity with continuous mode on the brane at
$z=0$. The boundary conditions of the potential applied to the wavefunction, Eq.~(\ref{Be1}), 
\be\label{Bound}
\psi_{m}\left(\frac{1}{k}\right)=0,\quad z>0;\quad \psi_{m}\left(-\frac{1}{k}\right)=0,\quad z<0
\ee allows us to determine how the parameter $\epsilon$ is related to the graviton masses, i.e.
\be\label{ENERGY}
m^2\sim \frac{\varepsilon^2 }{ 4} \,\left(\frac{J_{\nu}\left(\frac{m}{|k|}\right)}{J_{\nu+1}\left(\frac{m}{|k|}\right)}\right)^2.
\ee 

 For $|k|\ll1$, in the asymptotic regime the solution of the Eq.~(\ref{ENERGY}) is approximately given by
\be\label{ENERGY.2}
{\frac {2}{m\,\varepsilon}}=\tan \left( {\frac {m}{|k|}}-\frac{(2\,\nu+1)\pi }{4}\right)\quad\Rightarrow\quad {m}^{2}\cong\frac{\nu^2\,{\pi }^{2}\,{n}^{2}}{4},\quad n=1,2,3...\quad.
\ee The set of discrete states obtained, however, may be replaced by a continuous treatment for $|k|\ll1$, such that $\sum_{n}\rightarrow\int$. The correction to the four-dimensional Newtonian potential
generated by the massive modes, is given by second term of the Eq.(\ref{pot}).
Consequently, it is necessary to
divide this integral into two regions
\be\label{N2}
\delta V(r)\sim \frac{M_5^{-3}}{r}\left(\frac{ 2 }{|k|\,\left( \nu+1\right)^{2}}\int_0^{\frac{1}{ r_c}} { m\,\left( {\frac{m}{|k|}} \right) ^{2\,\nu}
\,{\rm e}^{-m\,r} }dm+\int_{\frac{1}{ r_c}}^{\infty}\, \frac{4\,{m}^{2}}{{\pi }\, \left( \varepsilon^2+ {4\,{m}^{2}} \right)}\,
 {{\rm e}^{-m\,r}}\,dm\right),
\ee
where we define the value $r_c=1/|k|$. In the first case, the general result is given in terms of the a confluent hypergeometric function of the first kind, such that
\ben\label{GEN}
\delta\,V(r)&\sim&\frac{{\rm e^{-\frac{r}{r_c}}}}{M^3_{5}\,r}\left(\frac{2}{{r_c}}\, 
\frac{{\mbox{$_1$F$_1$}(1;\,2\,\nu+3;\,{\frac {r}{r_{{c}}}})}}{\left( \nu+1 \right) ^{2} \left( 2\,\nu+2 \right)}
+\frac{1}{\pi\,r}\right)\nonumber \\ &+& {\frac {f(\nu)\, \left[ \sin \left( \frac{f(\nu)}{2}\,\frac{ r}{r_c}\right) {\rm Ci} \left(\frac{f(\nu)}{2}\,\frac{ r}{r_c} \right)
		-\cos \left( \frac{f(\nu)}{2}\,\frac{ r}{r_c} \right)\left( -\frac12\,\pi+{\rm Si} \left( \frac{f(\nu)}{2}\,\frac{ r}{r_c} \right) \right)\right] }{r_c\,r\,\pi\,M_5^3 }},
\een
where $f(\nu)=4+2\,\nu$. 
For $\alpha\sim {\rm O}\left(1/r_c\right)$ ($\nu=$ real integer number), the first term in (\ref{GEN})  generates the following class of potentials
\ben\label{RS}
\delta V (r) \sim\,\frac{r_c^{2\nu+1}}{\,M_5^3\, r^{2\nu+3}},
\een 
which reproduces the same behavior of Randall-Sundrum scenario \cite{RS1} as $\nu\to0$.
On the other hand, for the crossover scale being very large, i.e., $r_c\to\infty$ the second term in (\ref{GEN}) is dominant and
for small distances, i.e., $r/r_c\ll1$  we obtain the following form 
\be\label{4D.1}
V \left( r \right) \sim  \frac{ f(\nu)}{2\,r\,r_c{
M^{3}_{{5}}}\pi }
\,\left( \frac{f(\nu)}{2}\,\frac{ r}{r_c} \left( \gamma+\ln  \left( \frac{f(\nu)}{2}\,\frac{ r}{r_c}\right)  \right) -\frac{f(\nu)}{2}\,\frac{ r}{r_c}+\frac12\,\pi+{\cal O}(r^2)\right)
, 
\quad r<<{r_c}.
\ee
Notice that, at this limit  the potential has the correct $4D$ Newton's law with $1/r$
scaling.

Finally, for large distance, i.e., $r/r_c\gg1$, the second term in the potential Eq.~(\ref{GEN}) gives
 \be\label{5D.1}
V(r)\sim \,{\frac {2}{{r}^{2}\,{M^{3}_{{5}}}\,\pi }}
,\quad\quad r>>{r_c}, 
\ee
that recovers the laws of $5D$ gravity \cite{RGS,DGP}. 

\subsubsection{Second case: ``Barrier''$-$like potential}

\begin{figure}[ht]
\includegraphics[{height=05cm,width=07cm,angle=00}]{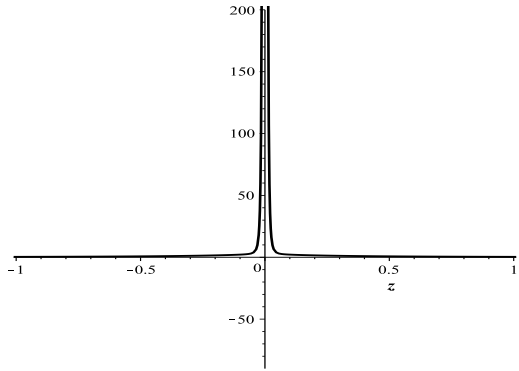}
\caption{Profile of the potential \eqref{pot.1} as a function of $z$ for $k>0$, or $\alpha^2=3/4\,k^2$.}\label{fig.2}
\end{figure}

Now, the analysis returns to the equation (\ref{sch.1}). In the second case, this a $\delta-$like potential problem, which come in two forms: for $k>0$ values in the 
potential (\ref{GPot}), or in the extreme case where $\frac{\alpha^2}{k^2}\sim\frac34$. Now, we have the reduced form $U(z)=3\,k\,\delta(z)$ in this case --- See Fig.~\ref{fig.2}.
%%%%%%%%%%%%%%%%%%%%%%%%%%%%%%%%%%%%%%%%%%%%%%%%%%%%%%%%%%%%%%%%%%
%%%%%%%%%%%%%%%%%%%%%%%%%%%%%%%%%%%%%%%%%%%%%%%%%%%%%%%%%%%%%%%%%%
The  probability density for the massive modes is given in terms of
the scattering states governed by $|{\psi}_{m} \left( 0 \right) |^2$ that depends on the magnitude of the transmission $T$ or reflection $R$
coefficients (as can be seen in the Ref.\cite{FBL2}). 
As usual, the jump condition at $z = 0$,
\be\label{JUMP1}
\left[{\frac {d}{dz}}{\psi}_{m} \left( z \right) \right
]_{{z=0}}=-3\,k\,{\psi}_{(m)} \left( 0 \right)
\ee
is obtained from the Schroedinger-like equation by using the properties of the delta function. We now consider the general
wave functions for scattered states in the form
\ben\label{Wave}
\psi_{1{m}} \left( z \right)& =&{{\rm e}^{i\,\kappa\,z}}+R\,{{\rm e}^{-i
\,\kappa\,z}},\,\,\quad z<0\nonumber\\
\psi_{{2m}}\left( z \right)&=&T\,{{\rm e}^{i\,\kappa\,z}},\quad\quad\quad\quad\quad z>0
\een
where ${\kappa}$ is the wave number.
The simplified form for the probability density  in this case is
%%%%%%%%%%%%%%%%%%%%%%%%%%%%%%%%%%%%%%%%%%%%%%%%%%%%%%%%%%%%%%%%%%%%%%%%%%%%%%%%%%%%%%%%
\ben\label{Prob1.1}
|T|^2=\left| {\psi}_{m} \left( 0 \right)  \right|^2={\frac {4\,{m}^{2}}{9\,k^2+4\,{m}^{2}}}.
\een 
The correction to the four-dimensional Newtonian potential
generated by the massive modes, is given by \cite{RS1}
\be\label{Corr}
V (r) = \frac{M^{-3}_{5}}{r}|\psi_0(0)|^2 + \delta V (r) ,
\ee 
Now, all the contribution to the Newton's law arises from the continuous massive modes on the brane such that we have
\be\label{int.1}
V(r)=\delta V(r)= \frac{M_5^{-3}}{r}
{\int_\frac{1}{r_c}^{\infty}} {   
  {\frac {4\,{m}^{2}}{9\,k^2+4\,{m}^{2}}}
\,{\rm e}^{-m\,r} }dm,
\ee where we define the crossover scale $r_c=1/|k|$. This is due to the fact that there is no zero mode in this case.
For the crossover scale being very large, i.e., $r_c\to\infty$, using the relation ${\rm Ei}(i\,x)={\rm Ci}(x)+i\left[-1/2\,\pi+{\rm Si}(x)\right]$, where ${\rm Ci}(x)=-\int_x^{\infty}{\cos\,t/t}$ and ${\rm Si}(x)=-\int_x^{\infty}{\sin\,t/t}$, the integral above turns
\ben\label{CiSik}
V \left( r \right) \sim {\frac {3}{2\,r\,r_c\,{M_{{5}}}^{3}}}\, \left[ \sin
 \left( \frac32\,\frac{r}{r_c} \right) {\rm{Ci}} \left( \frac32\,\frac{r}{r_c} \right) -\cos \left(\frac32\,\frac{r}{r_c} \right)
 \left( -\frac12\,\pi+{\rm{ Si}} \left( 
\frac32\,\frac{r}{r_c} \right)\right) \right], 
 \een which corresponds to the real part of the result of the integration (\ref{int.1}).
  For small distance, i.e., $r/r_c\ll1$  we obtain the following form 
\be\label{13k}
V \left( r \right) \sim \frac{3}{2\,r\,r_c{
M^{3}_{{5}}} }
\,\left( \frac32\,\frac{r}{r_c} \left( \gamma+\ln \left( \frac32\,\frac{r}{r_c}\right)  \right) -\frac32\,\frac{r}{r_c}+\frac12\,\pi+{\cal O}(r^2)\right)
, 
\quad r\ll{r_c}.
\ee
As we have found, in the previous cases, at short distances the potential has the correct $4D$ Newtonian $1/r$
scaling.

On the other hand, for large distance, i.e., $r/r_c\gg1$, the potential in Eq.~(\ref{CiSik}) gives
 \be\label{15}
V(r)\sim \,{\frac {1}{r^2\,{M_{{5}}^{3}}}}
,\quad\quad r\gg{r_c}, 
\ee
which naturally signalizes a $5D$ gravity behavior \cite{RGS,DGP}.
%%%%%%%%%%%%%%%%%%%%%%%%%%%%%%%%%%%%%%%%%%%%%%%%%%%%%%%%%%%%%%%%%%%%%%%%%%%%%%%%%%%%%%%%%%%%%%%%%%%%%%%%%%%%%%%%%%%%%%%%%%%%%%%%%%%%%%%%%%
%%%%%%%%%%%%%%%%%%%%%%%%%%%%%%%%%%%%%%%%%%%%%%%%%%%%%%%%%%%%%%%%%%%%%%%%%%%%%%%%%%%%%%%%%%%%%%%%%%%%%%%%%%%%%%
%%%%%%%%%%%%%%%%%%%%%%%%%%%%%%%%%%%%%%%%%%%%%%%%%%%%%%%%%%%%%%%%%%%%%%%%%%%%%%%%%%%%%%%%%%%%%%%%%%%%%%%%%%%%%%
\subsubsection{Third case: The extreme case where $\alpha^2<<k^2$ (or $\nu\sim\frac34$)}
The potential \eqref{GPot} now reduces to general form
\be\label{GPot.1}
U(z)=\frac{3}{4\,\left(1/k+|z|\right)^2}+3\,k\,\delta(z),
\ee  which is a specific case of the Bessel-like equations. The correction to the four-dimensional Newtonian potential
generated by the massive modes, is given by second term of the Eq.(\ref{pot}).
Consequently, it is necessary to
divide this integral into two regions
\be\label{N2.1}
\delta V(r)\sim \frac{M_5^{-3}}{r}\left(
\frac{ 32 }{49\,|k|}\int_0^{\frac{1}{ r_c}} {   
  m\,\left( {\frac{m}{|k|}} \right) ^{\frac{3}{2}}
\,{\rm e}^{-m\,r} }dm+\int_{\frac{1}{ r_c}}^{\infty}\, \frac{4\,{m}^{2}}{{\pi }\, \left( \varepsilon^2+ {4\,{m}^{2}} \right)}\,
 {{\rm e}^{-m\,r}}\,dm\right),
\ee 
where $\varepsilon=11/2\,|k|$ and we define the value $r_c=1/|k|$. For the crossover scale being very large, i.e., $r_c\to\infty$, using the relation ${\rm Ei}(i\,x)={\rm Ci}(x)+i\left[-1/2\,\pi+{\rm Si}(x)\right]$, where ${\rm Ci}(x)=-\int_x^{\infty}{\cos\,t/t}$ and ${\rm Si}(x)=-\int_x^{\infty}{\sin\,t/t}$, the integral above turns
\begin{eqnarray}\label{CiSik2}
\delta V \left( r \right) &\sim&-\frac{8}{49\,M_5^3}\,\left(\frac{49}{8}\,\frac{{\rm e}^{-\frac{r}{r_c}}}{\pi\,r^2}+\frac {4\,{\rm e}^{-\frac{r}{r_c}}}{r^2} +\frac{10\,r_c \,{\rm e}^{-\frac {r}{r_c}}}{r^3}+\frac{15\,r_c^2 \,{\rm e}^{-\frac {r}{r_c}}}{r^4}
-\frac{15\,\sqrt{\pi}\,{{r_c}^{5/2} }}{2\,r^{9/2}}\,{\rm erf}\left(\sqrt{\frac{r}{r_c}}\right)\right) \nonumber\\  
 &+&{\frac {11}{4\,\pi r\,r_c\,{M_{{5}}}^{3}}}\, \left[ \sin\left( \frac{11}{4}\,\frac{r}{r_c} \right) {\rm{Ci}} \left( \frac{11}{4}\,\frac{r}{r_c} \right) 
 -\cos \left(\frac{11}{4}\,\frac{r}{r_c} \right)\left( -\frac12\,\pi+{\rm{ Si}} \left( \frac{11}{4}\,\frac{r}{r_c} \right)\right) \right], 
\end{eqnarray} 
which corresponds to the real part of the result of the integration of the second term in (\ref{N2.1}).
{In this case, in addition to the discrete gravitational modes checked to $k<0$ (Eq. (\ref{ENERGY.2})), in the limit of the crossover scale is very small, i.e., $ r_c\to0$ (with the value $\nu\sim\frac34$ which means $\alpha\sim B_{23}\sim b^2\sim0$), the first term in (\ref{CiSik2}) is dominant and gives the following answer} 
\be\label{ERR}
V \left( r \right) \sim \frac{60\,\sqrt{\pi}\,{{r_c}^{5/2} }}{49\,{M^{3}_{{5}}}\,{r}^{9/2}},
\ee 
where we use the relations ${{\rm erf}\left(\sqrt {{x}}\right)}= {\frac {1}{\sqrt {\pi }}}\,\gamma \left( \frac12
,{x} \right)
$ and $\gamma\left(s,\frac{x}{y}\right)\to \Gamma(s)$, for $y<<x$.

On the other hand, for the crossover scale being very large, i.e., $r_c\to\infty$ the second term in (\ref{CiSik2}) is dominant, so for small distance, i.e., $r/r_c\ll1$  we obtain the following form 
\be\label{4D.4.2}
V \left( r \right) \sim  \frac{ {{11}}}{4\,r\,r_c{
M^{3}_{{5}}}\pi }
\,\left( \frac{11}{4}\,\frac{ r}{r_c} \left( \gamma+\ln  \left( \frac{11}{4}\,\frac{ r}{r_c}\right)  \right) -\frac{11}{4}\,\frac{ r}{r_c}+\frac12\,\pi+{\cal O}(r^2)\right), 
\quad r<<{r_c},
\ee and for large distance, i.e., $r/r_c\gg1$, the term in the potential Eq.~(\ref{CiSik2}) gives
 \be\label{5D.4.2}
V(r)\sim \,{\frac {1}{{r}^{2}\,{M^{3}_{{5}}}\,\pi }}
,\quad\quad r>>{r_c},
\ee that recovers the laws of $4D$ and $5D$ gravity, respectively.

\section{conclusions}\label{conclu}

In each case, in this paper, by considering massive graviton modes coming from a dimensional reduced brane solutions from supergravity, we have shown that the gravitational potential corresponds to the usual Newton potential which scales with $(\rightarrow 1/r)$ at short distance and has a five-dimensional behavior scaling with $(\rightarrow 1/r^2)$ at large distance compared with the crossover scale which depends on the parameter $k$ chosen in several ranges worked in the present analysis. In the regime of parameters considered here, we have a complete analysis of the emergence of gravitational modes where there is no regime of repulsive gravity.

\acknowledgments
The authors would like to thanks CNPq and CAPES  for partial support.

%%%%%%%%%%%%%%%%%%%%%%%%%%%%%%%%%%%%%%%%%%%%%%%%%%%%%%%%%%%%%%%

%%%%%%%%%%%%%%%%%%%%%%%%%%%%%%%%%%%%%%%%%%%%%%%%%%%%

\end{document}